# A computationally efficient queue-based algorithm for simulating volume-controlled drainage under the influence of gravity on volumetric images of porous materials

Jeff T Gostick[1,*], Niloofar Misaghian[1], Ashkan Irannezhad[2], Benzhong Zhao[2]

[1] Department of Chemical Engineering, University of Waterloo, Waterloo, ON, Canada

[2] Department of Civil Engineering, McMaster University, Hamilton, ON, Canada

[*] Corresponding author: jgostick@uwaterloo.ca

## Abstract

Simulating non-wetting fluid invasion in volumetric images of porous materials is of broad interest in applications as diverse as electrochemical devices and $CO_2$ sequestration. Among available methods, image-based algorithms offer much lower computational cost compared to direct numerical simulations. Recent work has extended image-based method to incorporate more physics such as gravity and volume-controlled invasion. The present work combines these two developments to develop an image-based invasion percolation algorithm that incorporates the effect of gravity. Additionally, the presented algorithm was developed using a priority queue algorithm to drastically reduce the computational cost of the simulation. The priority queue-based method was validated against previous image-based methods both with and without the effect of gravity, showing identical results. It was also shown that the new method provides a speedup of 20X over the previous image-based methods. Finally, comparison with experimental results at three Bond numbers showed that the model can predict the real invasion process with a high accuracy with and without gravitational effects.

# 1. Introduction

Volumetric images of porous materials can be obtained with resolutions as small as $2\ \mathrm{nm}$ per voxel [1], [2]. These images provide valuable information that can be used for various purposes in porous materials analysis and design. The types of structural information that can be obtained includes porosity, pore and particle size distribution, correlation length, and the degree of heterogeneity [3]–[5]. Additionally, images of porous materials can be used for numerical simulation of single-phase transport to find transport properties, such as permeability [6]–[9]. However, studying multiphase flow (the displacement of an immiscible fluid by another fluid phase) properties such as relative permeability in porous materials becomes more challenging as it is necessary to obtain the distribution of phases at the pore-scale. Studying multiphase flow in porous media has a wide range of applications, such as electrochemical devices [10]–[12] and $\mathrm{CO_2}$ sequestration [13], [14], so it is a high value target for improved techniques.

Although it is possible to capture images of fluid distributions in an experimental setup, such methods are more challenging than obtaining an image of a dry sample [15]–[17]. Specifically, it requires capturing multiple images at various saturations, segmenting multiple phases, aligning images, and in some cases accounting for changing wettability due to X-ray induced damage [18]. Moreover, capturing and postprocessing these images can be time-consuming, especially in large samples [19]. Due to these challenges, there is strong interest in developing numerical models to simulate multiphase transport at the pore scale. Techniques such as direct numerical simulations (DNS) provide detailed info on the phase distributions, as they solve the transport equations directly on realistic volumetric images of porous materials [20]–[22]. The Lattice Boltzmann Method (LBM) is commonly used for this but it is time consuming and memory intensive, and the problem grows exponentially with increasing image size [23]–[25].

A useful alternative method is the image-based approach, which involves inserting spheres into images as a proxy for the invasion of fluid-fluid interfaces. There are several ways to accomplish this, but the original method is based on morphological image opening (MIO) [26]–[28]. MIO-based methods have some limitations. Firstly, when simulating drainage, the saturation of the invading phase can experience large jumps, especially around the percolation threshold, and it is not possible to obtain fluid distributions at any intermediate saturations when such jumps occur. Secondly, the trapping of the defending phase is not captured properly, as the information between saturation jumps is lost. Thirdly, the effect of gravity is not inherently included in MIO-based approaches since they are based on geometrical constraints on capillary forces and do not include gravitational force. In our recent work [19] we presented an image based invasion percolation (IBIP) algorithm to simulate the invasion of non-wetting phase in a porous medium. Although that work addressed the first two problems in MIO-based approaches, the effect of gravity was not included. In another work, we developed an image-based drainage algorithm to include the effect of gravity, but it still has large saturation jumps [29]. The present work therefore combines our previous advances to create an accurate invasion percolation algorithm that incorporates gravitational effects.

The relative importance between gravity and capillarity is given by the Bond number, $\mathrm{Bo}$:

$$\mathrm{Bo} = \frac{\Delta\rho g R^2}{\sigma} \tag{1}$$

where $\Delta\rho$ is the density difference between the non-wetting and wetting phases, $g$ is the gravitation acceleration, $\sigma$ is the interfacial tension between the fluids, and $R$ is a characteristic pore size. Here, the median of the distance transform is used for $R$ [29]. For $Bo \ll 1$ gravity is expected to have negligible effect on the invasion process. As gravity becomes more important $(i.e., \mathrm{Bo} \to 1)$, the invasion front becomes more compact. The magnitude of gravitational effects on the invasion front also depends on the image size, though images are continually getting larger thanks to improving imaging technologies. Chadwick et al [29] explored the magnitude of errors that can arise as a function of image size and $Bo$, and found that gravity could be relevant in many cases of practical interest.

In addition, this study also presents a modified version of the Image-Based Invasion Percolation (IBIP) algorithm that substantially reduces the computational cost by using a *priority queue* to track the invasion sites rather than scanning the entire image to find the invasion sites at each step. The accuracy of the more efficient priority queue method was validated against the results of fluid configuration and pressure curves generated by previous image-based workflows without gravity [19] and with gravity [29]. Finally, gravity drainage experiments were performed in a micromodel at different Bo to further validate the developed framework. The code for the algorithms presented here is available freely in the open-source software package PoreSpy [30].

# 2. Methodology

In this section two separate algorithms are presented that incorporate gravity into image-based invasion percolation (IBIP) simulations. First, the original IBIP algorithm [19] is modified to incorporate gravity. Next, a novel, priority queue-based algorithm (QBIP) is presented. The QBIP algorithm produces equivalent results compared to IBIP but it is more than an order of magnitude faster.

## 2.1. Original IBIP Algorithm Using the Capillary Transform

The first algorithm presented here combines two concepts from previous works, allowing the inclusion of gravity effects [29] when performing image-based invasion percolation [31]. This is accomplished by recasting the invasion percolation problem in terms of a capillary pressure transform rather than a distance transform directly. In the new approach the invasion process operates by inserting spheres at the accessible voxel(s) with the lowest capillary pressure, which corresponds to the location(s) with the largest distance transform value. The capillary pressure transform, $T_{pc}$, is computed from the distance transform, $T_d$ as:

$$T_{pc} = \frac{2\sigma}{T_d \cdot L_{vx}} \quad (2)$$

where $\sigma$ is the interfacial tension of the fluid-fluid pair, $L_{vx}$ is the length of a voxel in unit of $\mathrm{m/voxel}$ and $T_d$ is the distance transform in unit of voxel. Note that the above expression assumes the invading fluid is perfectly non-wetting to the porous medium, which remains a limitation of sphere-based insertion methods, although some work has explored pathways to loosening this restriction [26], [32], [33]. Performing the invasion in terms of pressure instead of size is mathematically equivalent to the previous IBIP work, however, working with the capillary pressure transform provides a crucial benefit: it allows for the inclusion of gravity into the invasion process since the values in the capillary pressure transform can be adjusted to include elevation:

$$T_{pcg} = T_{pc} + \Delta\rho g h \quad (3)$$

where $\Delta\rho$ is the difference between the density of non-wetting phase and wetting phase, $g$ is the gravitational force, and $h$ is the height of the pixel relative to the datum in physical units. $h$ can be found as $h = z \cdot L_{vx}$ where $z$ is the height in unit of voxel. An image converted to a capillary pressure transform and adjusted by the gravity effect can thus be used as the input for the simulation. The algorithm proceeds following the flow diagram in Figure 1.

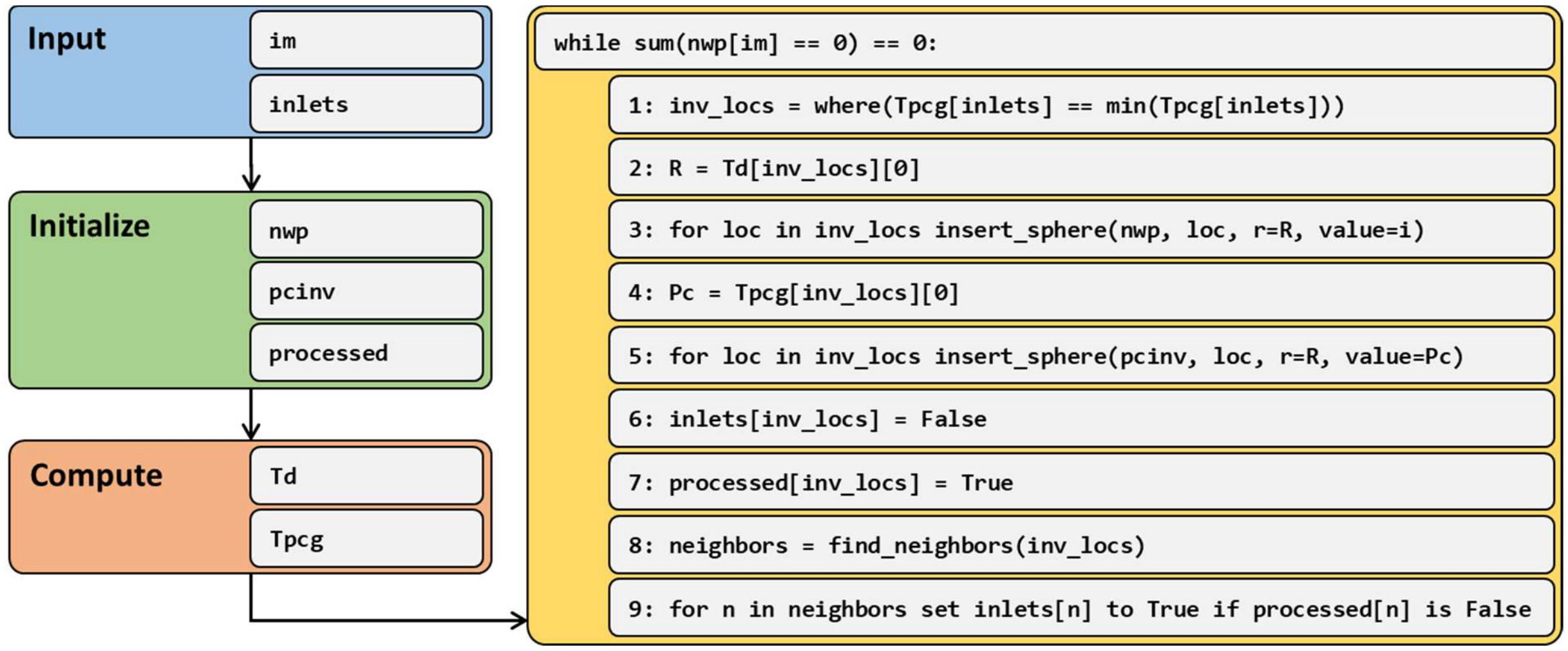

**Figure 1 Pseudocode flow diagram for IBIP algorithm operating on a capillary transform**

The only difference between the steps in Figure 1 and those in the previous work is the first step in the loop which finds the *minimum* $T_{pcg}$ at the locations of the inlets voxels instead of finding the *maximum* $T_d$. In the absence of gravity this modification is a straight-forward conversion between sizes and pressure using Eq.(2).

One challenge that arises when incorporating gravity into image-based drainage algorithms is the mismatch between the radius of curvature of the meniscus dictated by Eq.(3) and the radius of the sphere that must be drawn to fully occupy the pore. In other words, capillary pressure is impacted by both the size of the void space and the amount of gravity acting at the fluid-fluid interface. This means that spheres should be drawn using the radius indicated by the distance transform at each location, which results in menisci that do not have quite the correct curvature, meaning the menisci themselves are not impacted by gravity so are not flattened. This was discussed in detail by Chadwick et al. when analyzing the width of the invading front as a function of Bond number [29]. However, this only impacts the sharpness of the front, not the distribution of the non-wetting phase.

## 2.2. QBIP: Priority-Queue based Invasion Percolation Using the Capillary Transform

A key limitation of the original IBIP was slow speed, which resulted from performing repeated scans of the entire image to find the next invasion sites [19] (using Numpy's *where* function). Although it was shown that the computational cost can be reduced by employing GPU-based functions, the GPU showed only moderate advantage due to the bottleneck of transferring data from the CPU [19]. In addition, using GPU-based algorithms require special computational resources which are not always available, so they are not a panacea. Therefore, in this section, a new approach is presented which showed marked speed improvement. The main change was to replace the use of the *where* function with a *priority queue* that keeps an ordered list of which voxels are accessible for invasion and crucially, which should be invaded next. The new queue-based invasion percolation algorithm is referred to herein as QBIP. The priority queue has the useful property that the first item in the queue, known as the root, always has the highest priority (i.e., the lowest value capillary entry pressure) [26]. Therefore, the root element is always the next voxel to be invaded in each iteration so no searching or scanning needs to be performed. The main computational cost of priority queues is incurred when new items are added to the queue because they need to be added in a special order such that the root item has the highest priority. The *heapq* module in Python was used here, which is based on a binary heap. This data structure is relatively fast at inserting new elements, resulting in substantial reductions in processing time. Moreover, the *heapq* module is compatible with the just-in-time compilation library *numba*, which was essential to obtain the observed speed-up. It should be noted that the algorithm proposed here is agnostic to the underlying heap so any compatible implementation could be used (e.g. the Fibonacci heap) if desired. The developed workflow is shown in Figure 2. The steps shaded grey differ from the workflow shown in Figure 1.

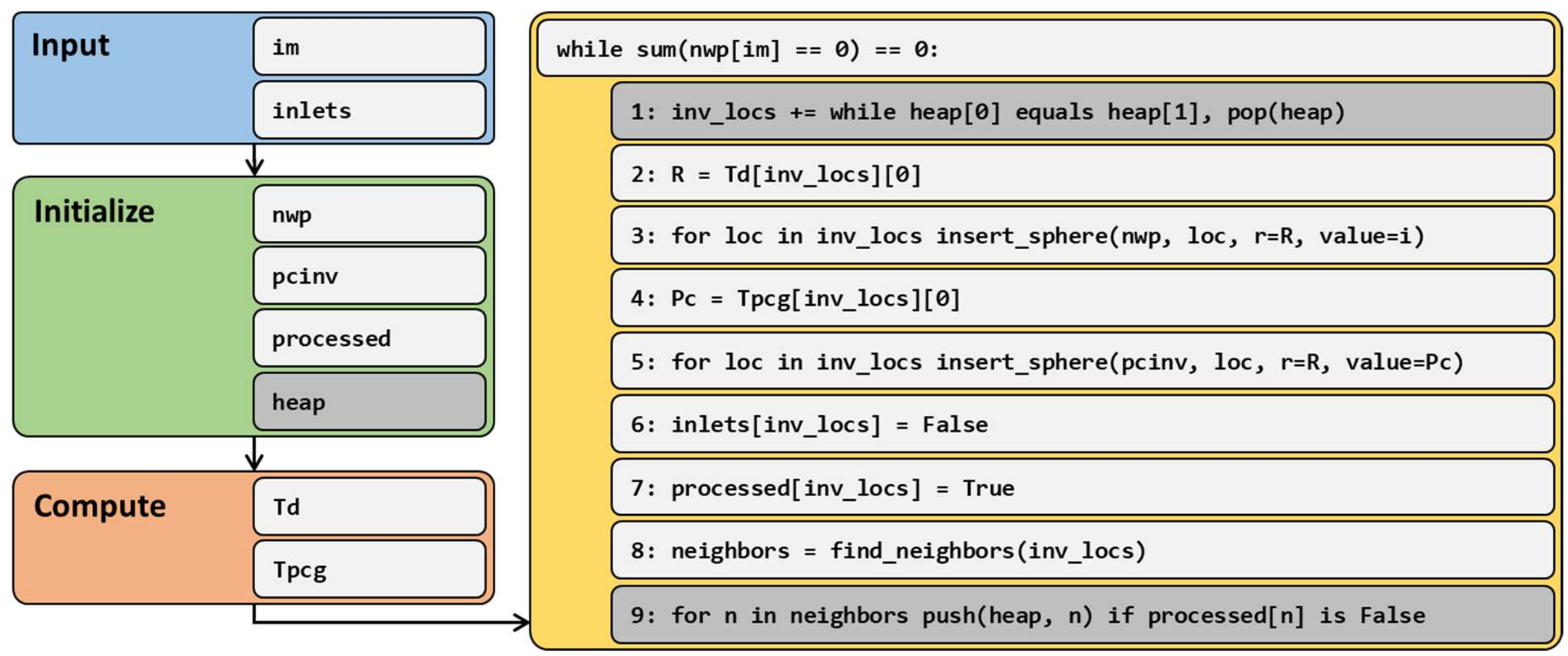

**Figure 2: Pseudocode flow diagram for the IBIP algorithm using a priority queue and the capillary transform**

The algorithm starts with similar steps as explained in Section 2.1 to compute the capillary pressure transform. Next, instead of processing the image directly, a list of values is created where each row contains the value of $T_{pc}$, $T_d$, and coordinates of pixels in the *inlets* image. The list is then converted to a binary heap with capillary pressure values used to determine priority. The method begins by *popping* the root element from the heap to find the first invasion point.

However, since there could be multiple sites with the same invasion pressure, the new root is inspected and popped until a root with a higher invasion pressure is found. This results in a list of one or more voxels where invasion should occur simultaneously. As the items in the queue have both the $T_d$ and coordinates of the voxels, this information is used to insert spheres with the radius taken from the $T_d$. The spheres are inserted into $nwp$, writing values of the current step number, and into $pcinv$ writing capillary pressure values. Next, the uninvaded neighbouring voxels of those which were invaded are found and *pushed* onto the queue. The process of pushing onto the queue ensures that the items are placed at the correct location according to their respective capillary pressures. The process continues until a maximum number of iterations is reached or the void space in the $nwp$ image is filled, whichever comes first. A speed comparison between the QBIP approach and the previous IBIP algorithm is shown in Figure 3, and a speedup of more than 20x is evident, lowering computational time from hours to minutes for a given image. The QBIP algorithm was used for the remainder of the present manuscript due to its computational advantages, though it should be stressed that both the IBIP and QBIP algorithms give identical results, and both can be modified to incorporate gravitational effects. It was confirmed that the IBIP and QBIP algorithms output exactly the same results with pixel-perfect agreement. This comparison is provided in the supplemental information.

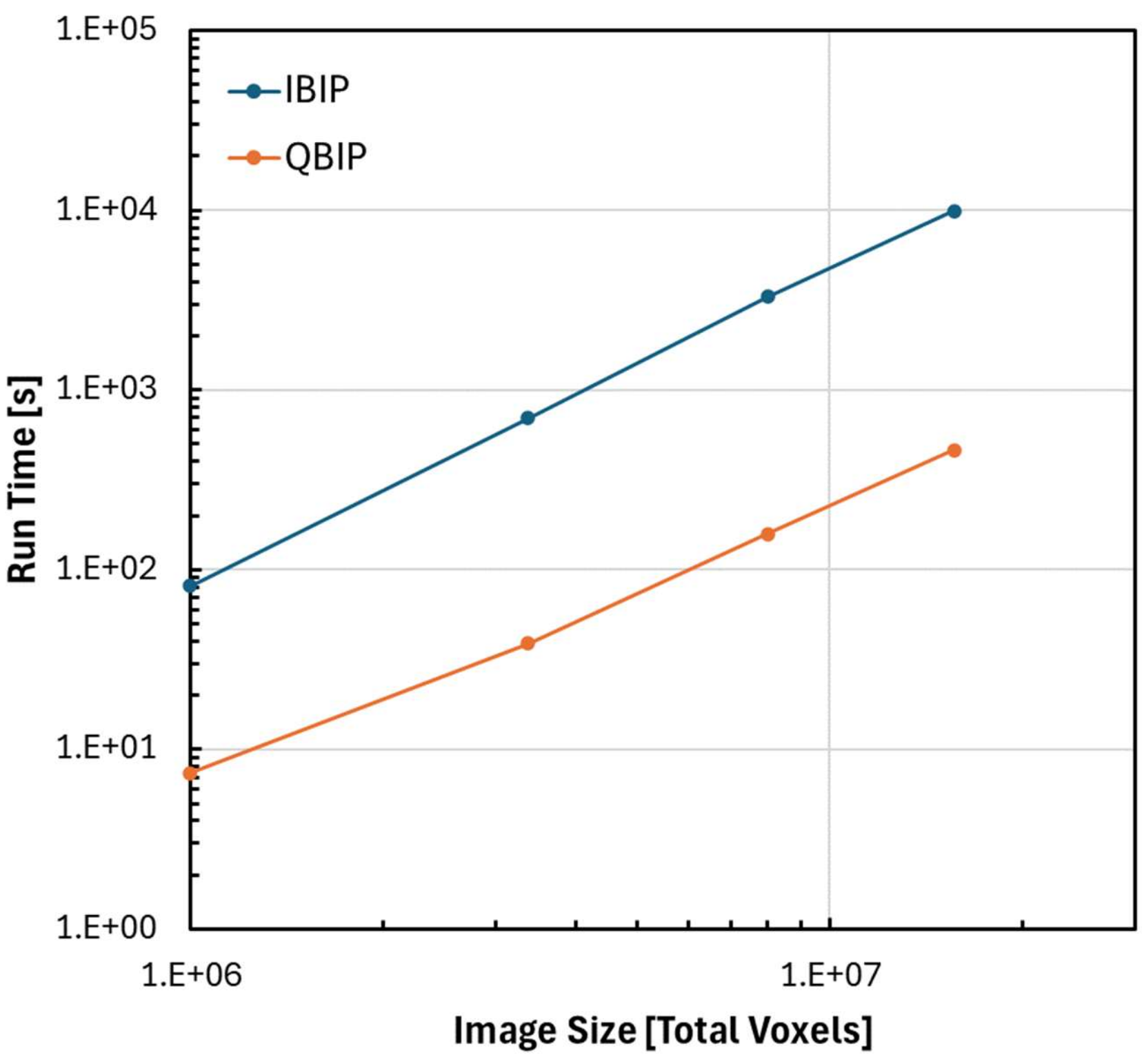

**Figure 3: Speed comparison between the original IBIP algorithm, and the new version presented here based on a priority queue.**

## 2.3. Identifying Trapped Voxels

It is well-known that invasion percolation of a non-wetting fluid in porous media results in the trapping of the defending wetting fluid. In principle, identifying trapped voxels in image-based invasion algorithms is a straightforward postprocessing step applied to an invasion sequence map. For instance, in the simple case where the outlet is one voxel, any voxels with a higher invasion sequence value than the outlet must have been invaded *after* the non-wetting fluid has already reached the outlet, so those voxels are all trapped. The logic is similar in the more realistic case where the outlets encompass the entire face or edge of an image, though the process must be performed for all invasion sequence values at the outlets. This can be accomplished by finding all voxel clusters whose sequence value is less than or equal to $N$, then marking all voxels not connected to the outlets as trapped. The process is repeated for the next higher value of $N$. If the invading fluid configuration were determined using IBIP then the maximum number of times this needs to be performed will not exceed the number of pressure steps applied, which is usually less than 100 (i.e. 100 steps between 0 and $P_{c,max}$). In QBIP, however, this approach can become unreasonably time consuming since most voxels have their own unique invasion sequence value (so on a 200-cubed image with 50% porosity this means $\sim 10{,}000$ cluster searches). Therefore, a new algorithm was developed which also happens to use a priority queue so has approximately the same computational cost as the QBIP algorithm itself, otherwise the computational gains of the QBIP algorithm would be of less practical use. The new queue-based trapping algorithm was inspired Masson [34] who used a similar approach to find trapped pores in a pore network model. In the present case, an invasion percolation process is run "in reverse" with priority determined by the higher invasion sequence number. Initially the outlet voxels are added to the queue, then it proceeds by popping the root node (repeatedly until the next root has a smaller value). The popped voxel(s) is marked as trapped if it has a higher value than the lowest value seen thus far. All the neighbors of the popped voxel(s) are added to the queue (if they aren't already). The process continues until all the voxels have been visited. This allows for the invasion front to recede from the outlets in the same order it reached the outlets but bypassing any trapped voxels. Note that once a voxel is marked invaded there is no need to draw spheres as is done in the QBIP algorithm.

One problematic aspect of both the cluster-based and the queue-based trapping algorithms is that small clusters of isolated voxels on the solid surfaces can become erroneously trapped due to the digitized nature of the image. This is illustrated in Figure 4. Panel (a) shows the invasion sequence map without accounting for trapping, while panel (b) shows the result after applying trapping with black indicating trapped pixels. (Note that the cluster-based and queue-based algorithms both produce identical outputs, as shown in the supplemental information). The large region in the upper left corner was correctly identified, but several small clusters of trapped pixels can be seen along the edges of the solid which are clearly erroneously identified. An additional step was therefore necessary to return these voxels back to "untrapped". This was done by finding all clusters of trapped voxels smaller than some threshold size (10 was used in the present work), then removing them from the mask of trapped voxels. The sequence values for these voxels were taken as the minimum of all the non-zero sequence values that border the cluster.

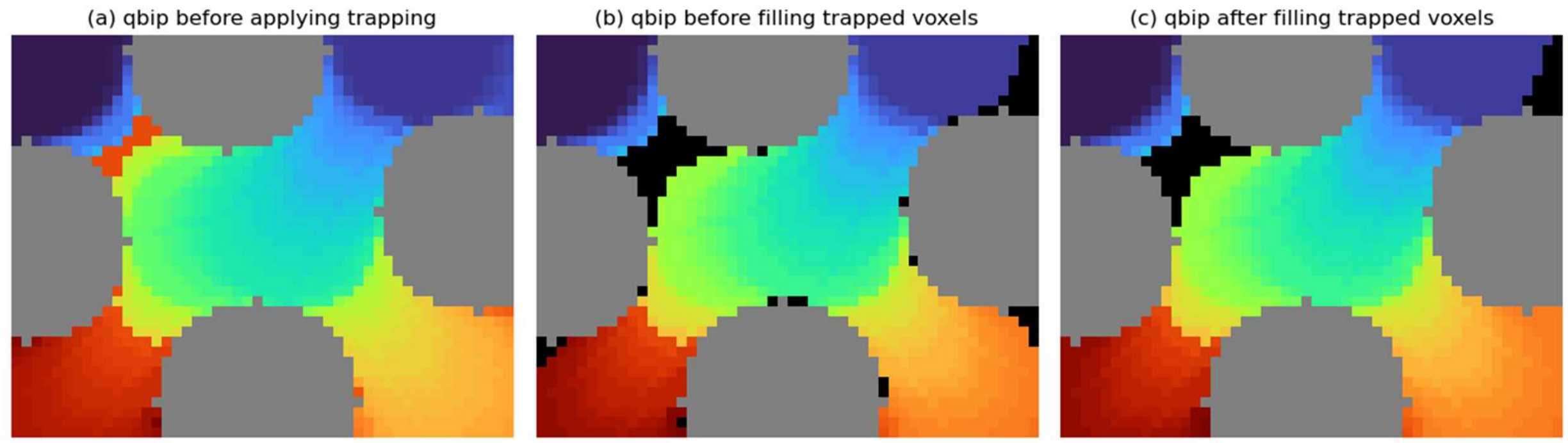

**Figure 4: Distribution of non-wetting phase before (a) and after (b) applying the trapping algorithms, and (c) after filling trapped voxels at solid surfaces. Pixel color corresponds to invasion step, with grey being solid and black indicating trapped pixels.**

## 2.4. Micromodel Experiments

To validate the QBIP algorithm, gravity drainage experiments were conducted in the micromodel shown in Figure 5a. The micromodel consists of two acrylic sheets of $10\ \mathrm{cm} \times 60\ \mathrm{cm}$: a flat sheet and a patterned sheet that are held together by rows of bolts. The pattern contained more than 8,000 laser engraved cylindrical posts (Figure 5a). The height of the posts is 500 µm, which also sets the gap thickness where fluid flow takes place. The post pattern in the micromodel was designed by first generating an irregular triangular mesh using MATLAB's *pdemesh* tool. The nodes of the triangular mesh serve as the centers of the posts. Each post was then assigned a radius of 45% of its nearest neighbor's distance, resulting in a random distribution of non-overlapping posts. This pattern contains local disorder, but it is macroscopically homogeneous at the scale of micromodel. It should be noted that three sides of the flow cell are impermeable walls, while only the top side is open to the air. The throat sizes and solid post sizes are shown in Figure 5b.

To perform an experiment, the cell was first fully saturated with silicone oil ($\mu_{oil} = 10\ \mathrm{mPa \cdot s}$, MilliporeSigma, USA) dyed with $600\ \mathrm{mg/L}$ Oil Blue N (MilliporeSigma, USA). During an experiment, silicone oil was withdrawn from a port at the bottom of the micromodel at a constant volumetric rate $Q = 0.0153\ \mathrm{mL/min}$, which was accompanied by air invasion from the open top boundary of the micromodel. The capillary number associated with the drainage process was $Ca = 1.82 \times 10^{-4}$. The defending fluid (i.e., silicone oil) completely wets the acrylic surface in the presence of air [14], [35]. The micromodel was illuminated by an LED light panel and the invasion process was captured by a DSLR camera (D850, Nikon, Japan). The image resolution of the experiments is $74\ \mathrm{\mu m}/pixel$. A series of image-processing techniques including contrast enhancement, perspective correction, background subtraction, and thresholding were applied on the raw images to segment the images into three labels (i.e., posts, air, and silicon oil). The segmented images were used for comparison with modeling results, which will be discussed later. The experiments were performed at angles of 0 °, 30 °, and 90 ° degrees relative to the horizontal plane. The corresponding Bond numbers of each setup were calculated using Eq.(1) resulting in values of $\mathrm{Bo} = 0$, $0.05$, $0.1$, respectively.

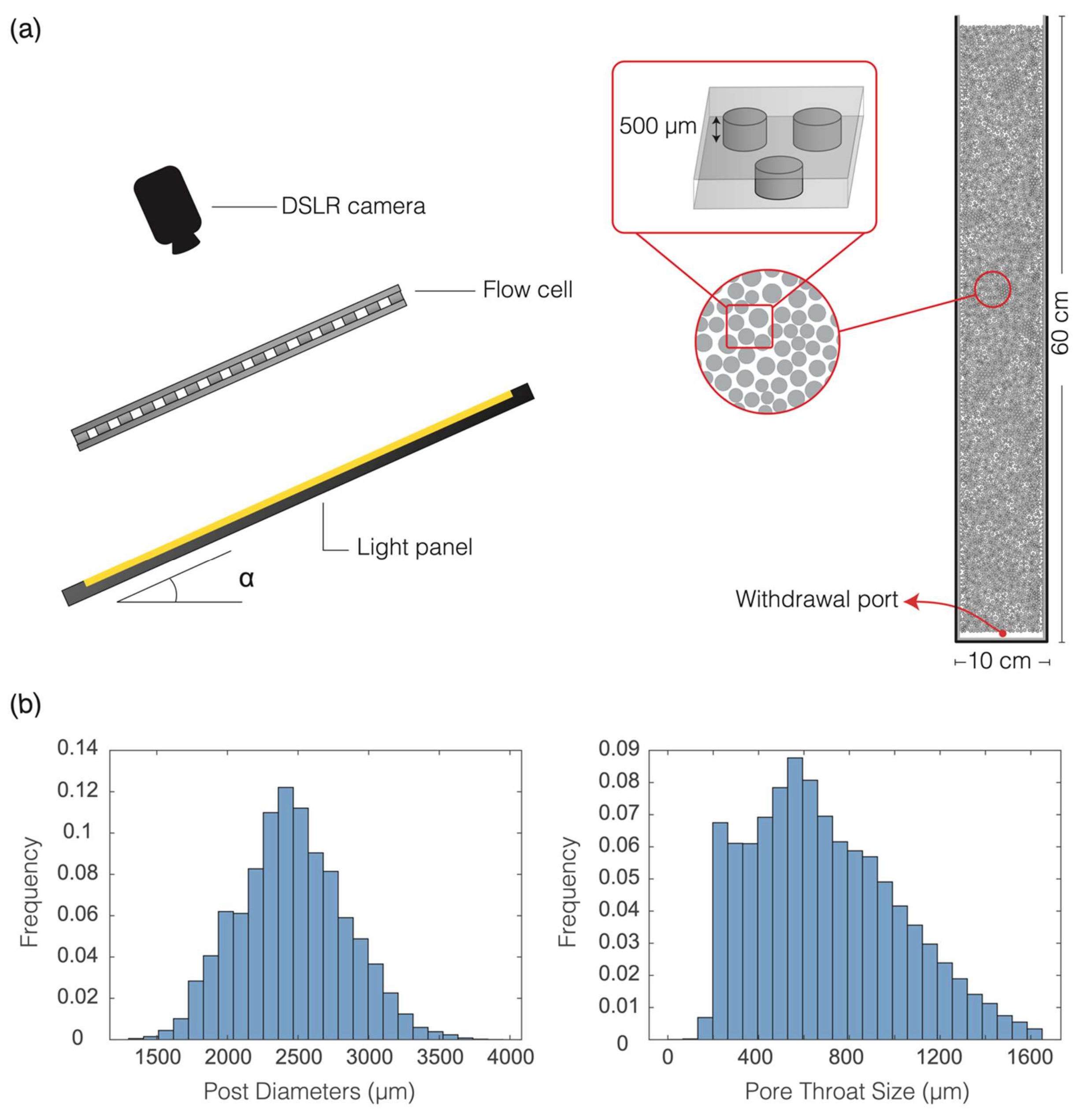

**Figure 5 (a) Gravity drainage experiments were conducted by withdrawing viscous silicone oil from a quasi-2D micromodel patterned with cylindrical posts. The height of the posts was 500 µm. The post pattern is designed to introduce local disorder, but the micromodel is macroscopically homogeneous. (b) The post diameters range from 1500 to 3700 µm (left), while the pore-throat sizes range from 100 to 1600 µm (right).**

# 3. Results

## 3.1. Incorporating gravity into QBIP and Validation

For initial validation purposes, the gravity drainage algorithm by Chadwick *et al.*[29] was used as a reference. That algorithm applied increasing pressure steps to drive non-wetting phase invasion, which is an ordinary percolation process. Herein this will be referred to as image-based ordinary percolation (IBOP). Figure 6 compares the output of the QBIP algorithm with the results of the standard IBOP simulation in the absence of gravity. Figure 6(a) and (b) show the saturation maps, while Figure 6(c) shows the drainage capillary pressure curves. The fluctuating capillary pressure readings of the QBIP result is expected since it is effectively a volume-controlled invasion, meaning the pressure fluctuates as the invasion front squeezes through different size constrictions in the void space. Indeed, these pressure fluctuations have been observed experimentally by Måløy et al [36]. Figure 7 shows the same results as Figure 6 but for the case where gravity cannot be neglected ($\mathrm{Bo} = 0.1$). The saturation maps in Figure 7(a) and (b) show a more compact invasion front, as is expected in gravity stabilized displacements. The capillary pressure curve in Figure 7(c) is noticeably sloped which is also expected in the presence of gravity. This curve is technically a “pseudo-capillary pressure curve” since the pressure in the dense phase varies throughout the domain due to the static pressure differences at different elevations throughout the fluid. The reported capillary pressure is with respect to a datum, which in this case is the bottom of the domain. The increasing sharpness of the displacement front is confirmed visually in Figure 8. Each panel corresponds to an increasing value of $\mathrm{Bo}$, and each line is the saturation profile at global non-wetting phase saturation values. Figure 8 contains both the QBIP results as the line and the IBOP results as the circular markers, and the agreement between the two algorithms is essentially perfect. Note that because the IBOP algorithm experiences large jumps in saturation between successive pressure steps, the lines are each computed at different global saturations corresponding to a specific fluid composition in the IBOP results. The main takeaway from this section is that gravity can be easily incorporated into the QBIP algorithm by using Eq.(3) to compute the capillary pressure transform, and that excellent agreement can be obtained with other established methods. In the next section the output of the QBIP algorithm will be compared to micromodel experiments.

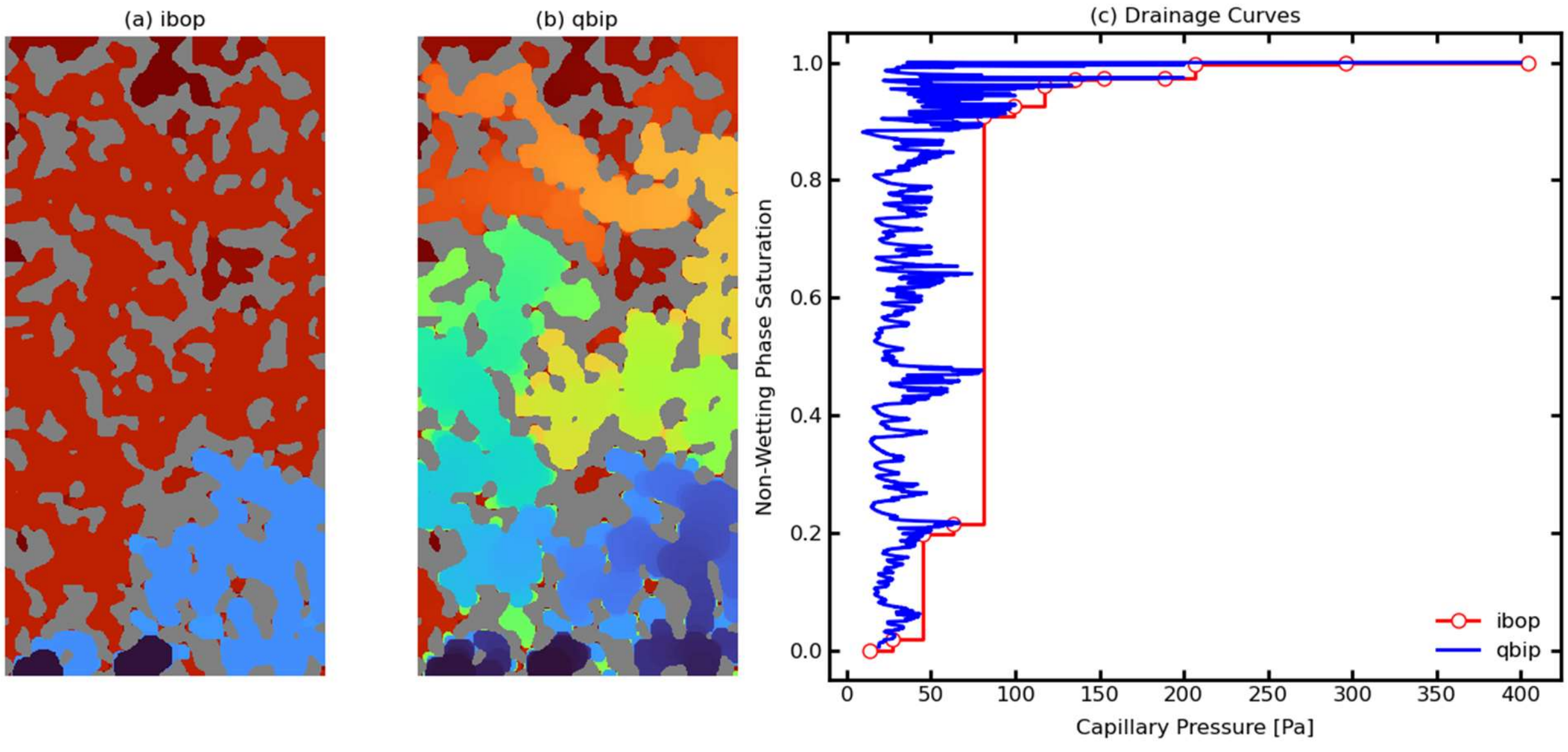

**Figure 6: Saturation maps for IBOP (a) and QBIP (b) along with the drainage capillary pressure curve for both (c) with $Bo \ll 1$. The color map corresponds to the invasion sequence with grey pixels being solid phase.**

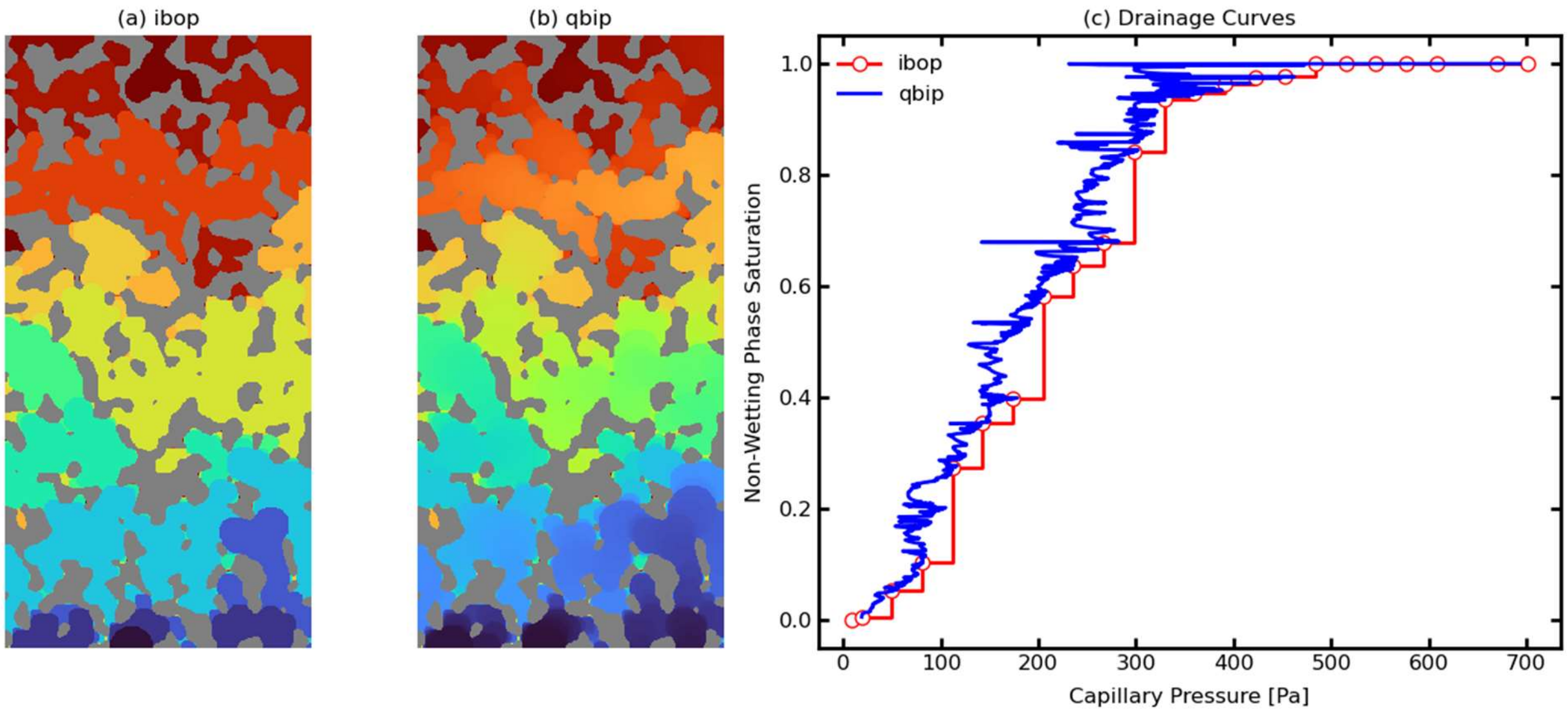

**Figure 7: Saturation maps for IBOP (a) and QBIP (b) along with the drainage capillary pressure curve for both (c) with $Bo = 0.1$. The color map corresponds to the invasion sequence with grey pixels being solid phase.**

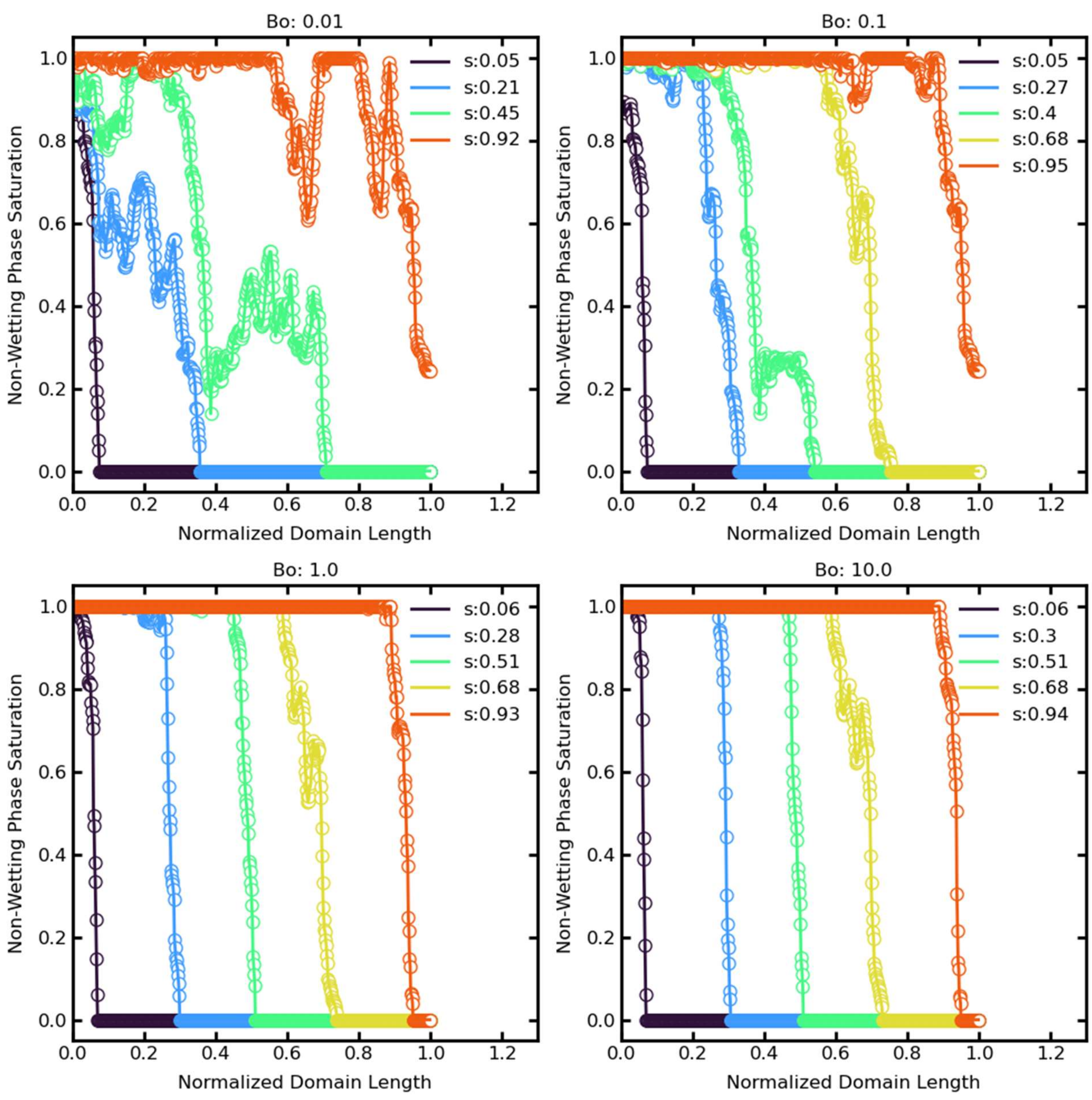

**Figure 8: Saturation profiles through domain as a function of $Bo$. Circular markers are the IBOP results and the lines are the QBIP results, showing perfect agreement.**

## 3.2. Comparison to Micromodel Experiments

The micromodel results are shown in Figure 9(a-c) for the horizontal experiment (α=0°), inclined experiment (α=30°), and vertical experiment (α=90°), respectively. To facilitate comparison, 9 image sequences taken at different saturations were combined into a single saturation map (equivalent to what is produced by the QBIP algorithm). The cluster-based trapping algorithm described in Section 2.3 was also applied to the final image to identify any clusters of wetting phase which were trapped at step $N$ but disappeared from the images as step $> N$ which was attributed to a combination of evaporation or film flow, neither of which were included in the simulation. In other words, wetting fluid that was trapped in image sequence 4 might have disappeared by image sequence 5 or 6, so the physically trapped fluid in the final image does not reflect the actual trapping that occurred. Since these additional physics were not included in the QBIP algorithm the experimental images were corrected accordingly.

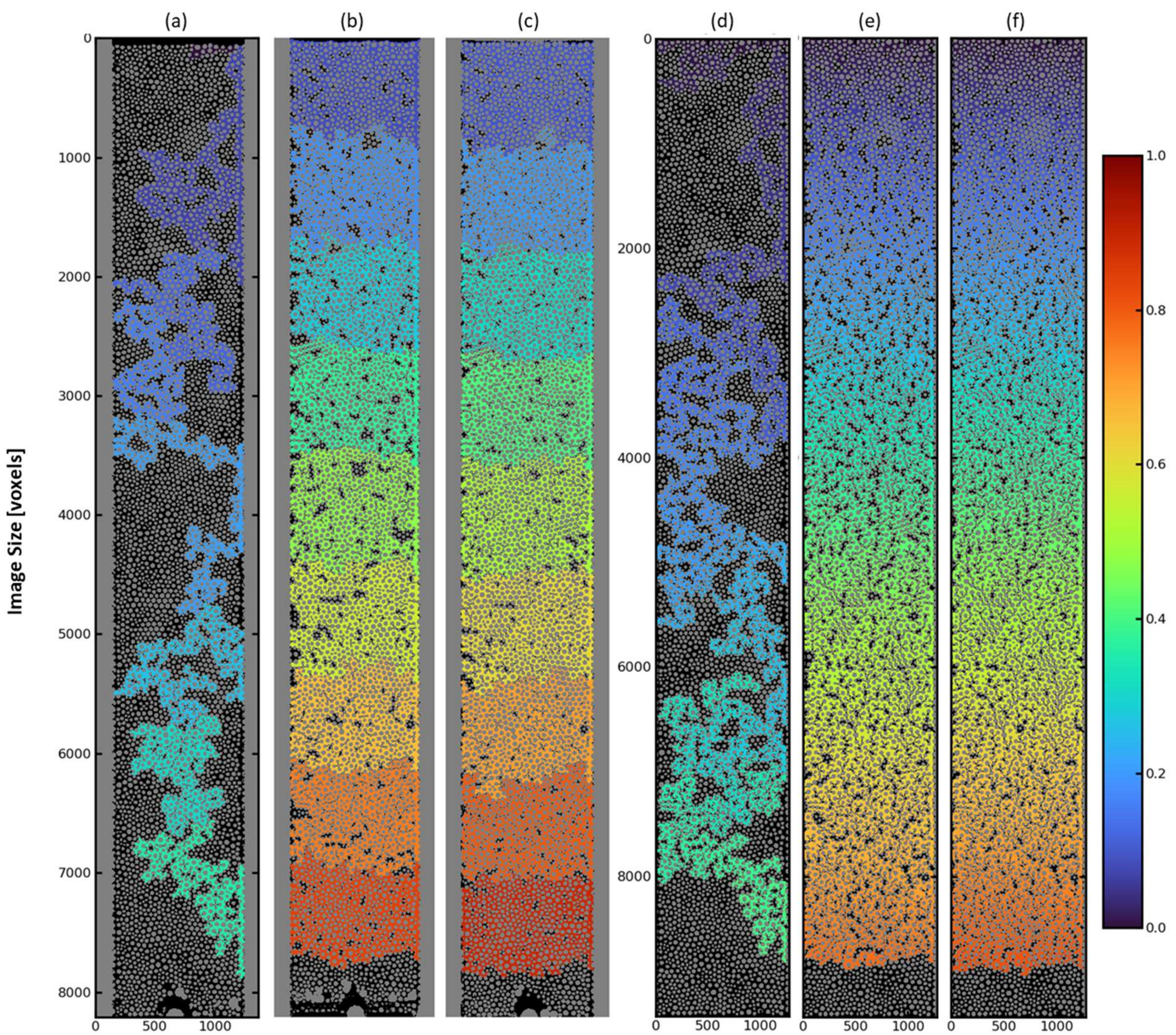

**Figure 9: Sequence maps of micromodel experiments at 0° (a), 30° (b) and 90° (c) inclination angles, and QBIP simulations for the same cases are shown in (d, e and f) respectively. Grey indicates solid, black is uninvaded**

**void space, including trapped wetting phase. The color and color-bar correspond to the global saturation in the domain at the point where the voxel was invaded.**

It should be noted that the QBIP simulations were performed on the CAD drawing of the micromodel design instead of the experimental image for two reasons: (i.) Lens distortion in the imaging process can lead to unphysical stretching of certain pixels; (ii.) Laser etching of the posts has a tolerance on the order of 100s of micrometers. Taking these into account, satisfactory agreement between the experiment and simulation is achieved by the QBIP algorithm.

Two changes were required to simulate the experiments. Firstly, since the micromodel is a quasi-2D porous medium with an out-of-plane dimension (s), the capillary pressure transform of the image was calculated using the following equation instead of Eq.(2):

$$P_{c_i} = \sigma \left( \frac{1}{T_d L_{vx}} + \frac{2}{s} \right) \tag{4}$$

Note that a 3D image could have been generated by extruding the 2D image, but the approach given by Eq.(4) is actually more accurate since it includes both radii of curvature in the capillary pressure transform, while the 3D image would use the same distance transform value for both directions. Secondly, the density difference between the two fluids was calculated using the same convention ($\rho_{nwp} - \rho_{wp}$), but because the density of the invading phase (air) was lower than the defending phase (silicone oil), a negative value of $\Delta\rho$ was used in Eq.(3). The QBIP algorithm is able to handle this without any numerical issues, but the non-wetting fluid must enter the domain from the top to maintain the gravity stabilized configuration, which is of course also true of the experiment. It should be noted that the QBIP algorithm can technically perform simulations in the gravity destabilized direction, though it is left for future work to evaluate whether this provides a realistic invasion pattern. For instance, in its present form the QBIP algorithm does not incorporate snap-off of the non-wetting phase, or re-imbibition of the wetting phase behind the invasion front. After considering all the above adjustments and caveats, the non-wetting fluid distributions predicted by the QBIP algorithm are shown in Figure 9(d-f). The compactness of the invasion front is evident for both the $\mathrm{Bo} = 0.05$ and $\mathrm{Bo} = 0.1$ cases.

Figure 10 compares the saturation profiles throughout the domain for the experimental (top row) and simulated (bottom row) snapshots for all Bo numbers studied here.

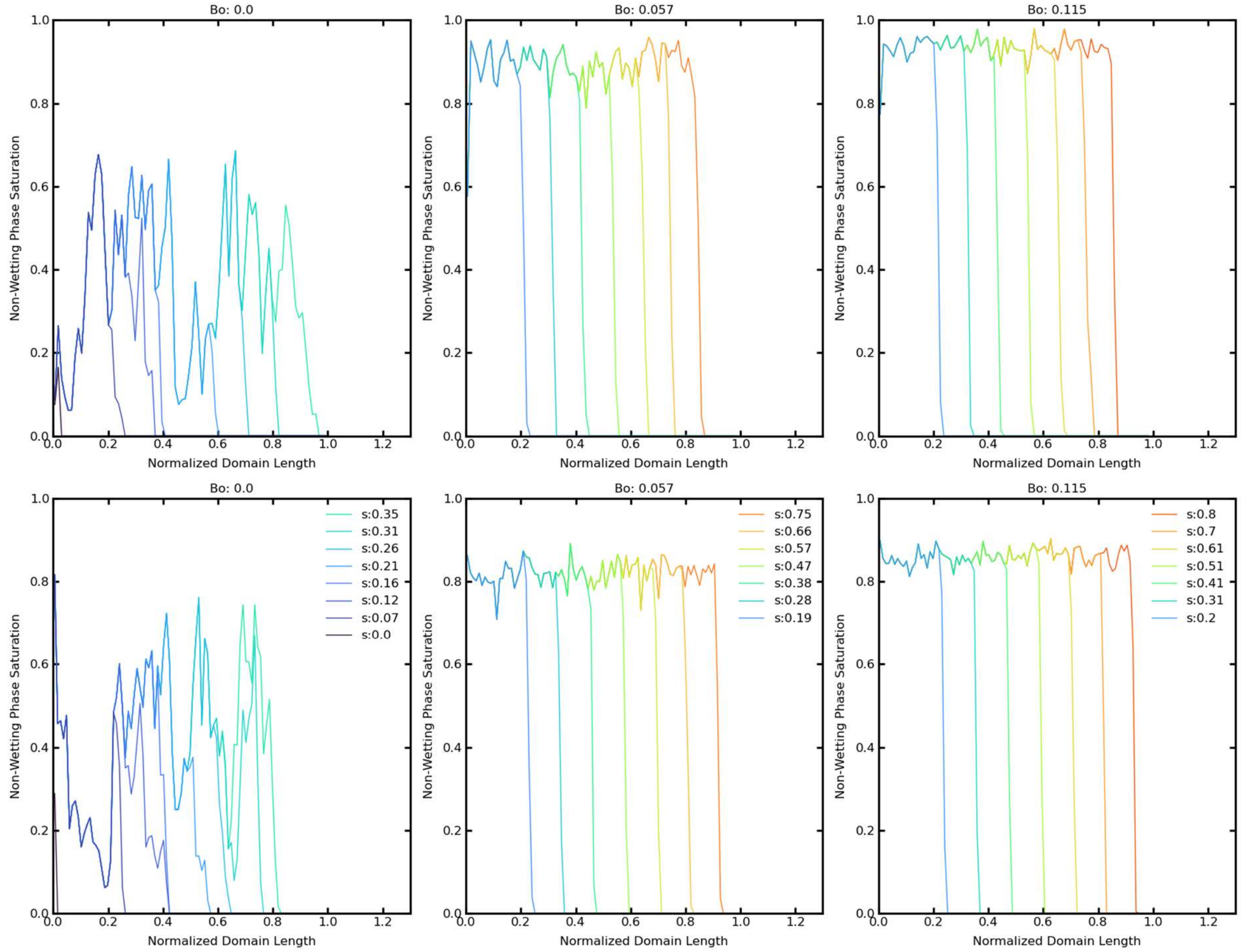

**Figure 10: Saturation profiles as a function of domain length. The top row represents experimental measurements, while the bottom row represents simulation results. Th columns represent horizontal, inclined, and vertical cases, corresponding to $\mathbf{Bo} = \mathbf{0.0}, \mathbf{0.05}, \mathbf{0.1}$, respectively.**

Qualitatively the invasion fronts appear to be in excellent agreement, and this is confirmed in Figure 11 which plots the normalized domain position of the leading edge of the front as a function of saturation for both the experimental and simulated results.  Clearly there is very good agreement for all cases over the full range of non-wetting phase saturation. For the case of $\mathrm{Bo} = 0$, the experimental data show the wetting front to be slightly more advanced for a given saturation. This could be attributed to minor viscous effects in the experiments due to the non-zero imposed flow rate. For the cases with $\mathrm{Bo} > 0$, the QBIP algorithm shows a lower saturation for a given height, which can be attributed to the larger amount of wetting phase trapping that is observed in the simulations compared to the experiments. Overall, the good agreement between the experiment and simulation results supports the accuracy of the QBIP algorithm.

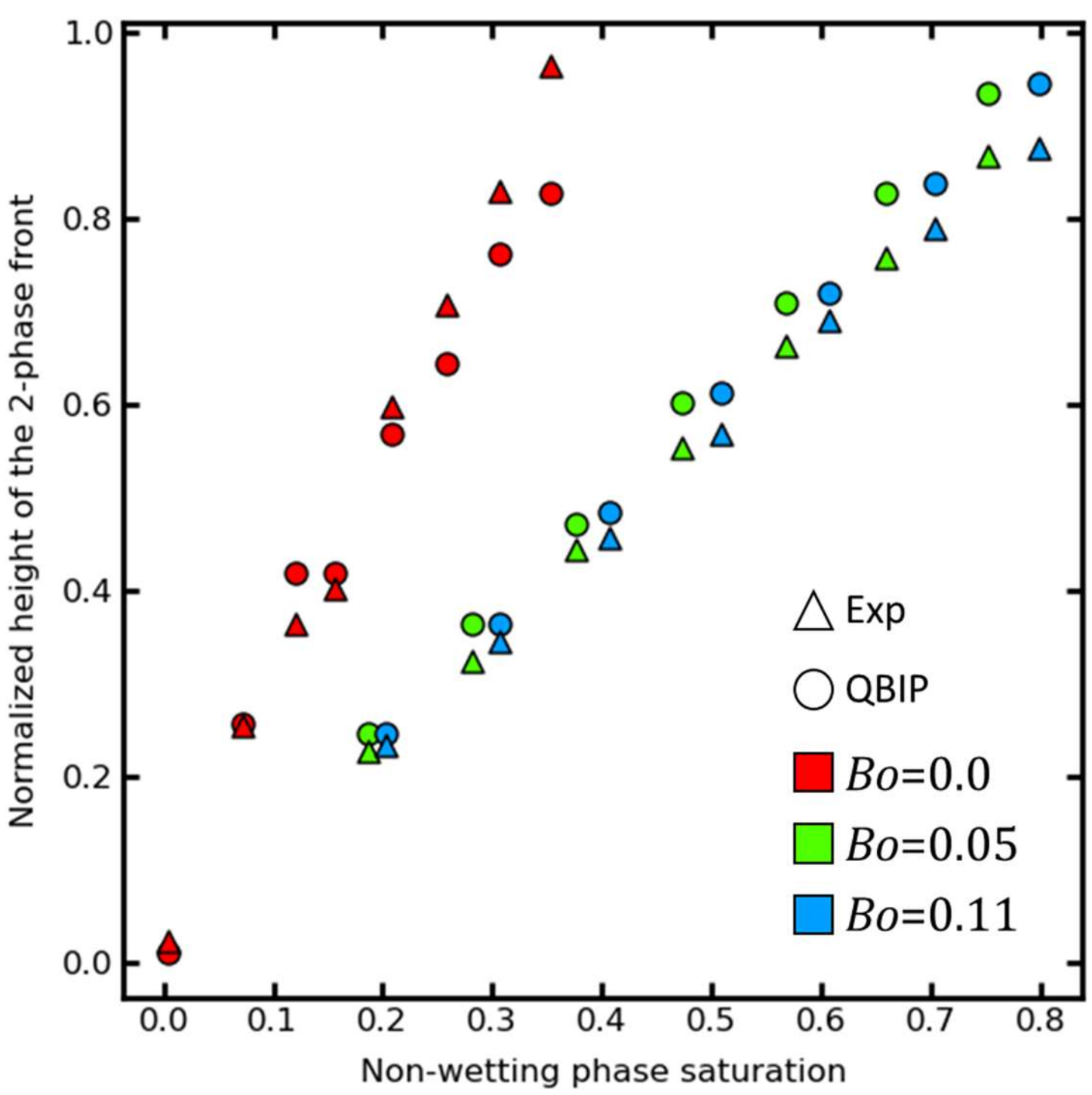

**Figure 11: Height of the invasion front vs global saturation for different Bo numbers.**

## 3.3. Resolution Dependence

All image-based algorithms are sensitive to the image resolution. The IBIP and QBIP algorithms are affected by resolution in two ways. Firstly, the distance transform is sensitive to the image resolution since these values are computed relative to the pixelated perimeters of the solid grains in the porous media. This issue is common to all image analysis tools, but as will be discussed next has an oversized implication on invasion simulations. The second issue is unique to the invasion percolation process due to the pixel-by-pixel (or voxel-by-voxel) progression of the IBIP and QBIP algorithms. Consider the scenario where multiple throats (e.g. A and B) with similar sizes are accessible at the same time, both algorithms will invade a throat A, and only once all the void space that lies beyond that throat is filled will it invade throat B. If the image resolution changes it may happen that the order of invasion is swapped so throat B is invaded first (due to the pixelated nature of the distance transform). In this case a different portion of the image will be invaded first (that which is connected to throat B). This can lead to notably different invasion patterns, depending on the nature of the image. In the case of the micromodels discussed in the previous section, the results can be visibly different as seen in Figure 12.

Another way to look at this issue is by plotting the capillary pressure curves for different resolutions. This is illustrated in Figure 13, where QBIP was performed on an image of randomly located, non-overlapping spheres. The resolution of the image was decreased using a zoom factor < 1, and capillary pressure curves were plotted for each case. The blue curves show the pressure fluctuations of the invasion process as fluid fills different regions of the domain. At first glance the curves appear similar, but closer inspection reveals that some peaks have moved. For instance, the green and red arrows point to two separate peaks, which represent the filling of specific areas of the domain. In panel (c) the peak indicated by the red arrow moves dramatically, indicating that this region is filled much earlier in the process than in panel (b). In panel (d) the red arrow returns to its original location, but the green arrow is now filled at the start of the process. In panels (e) and (f) the peaks return to their original locations. This explicitly illustrates the sensitivity of the QBIP (and IBIP) algorithms to image resolution. Note however that this case is a bit contrived since the domain is quite small, with a relatively small number of filling events (i.e. peaks). On a larger image the movement of these peaks would still occur but would be less prevalent.

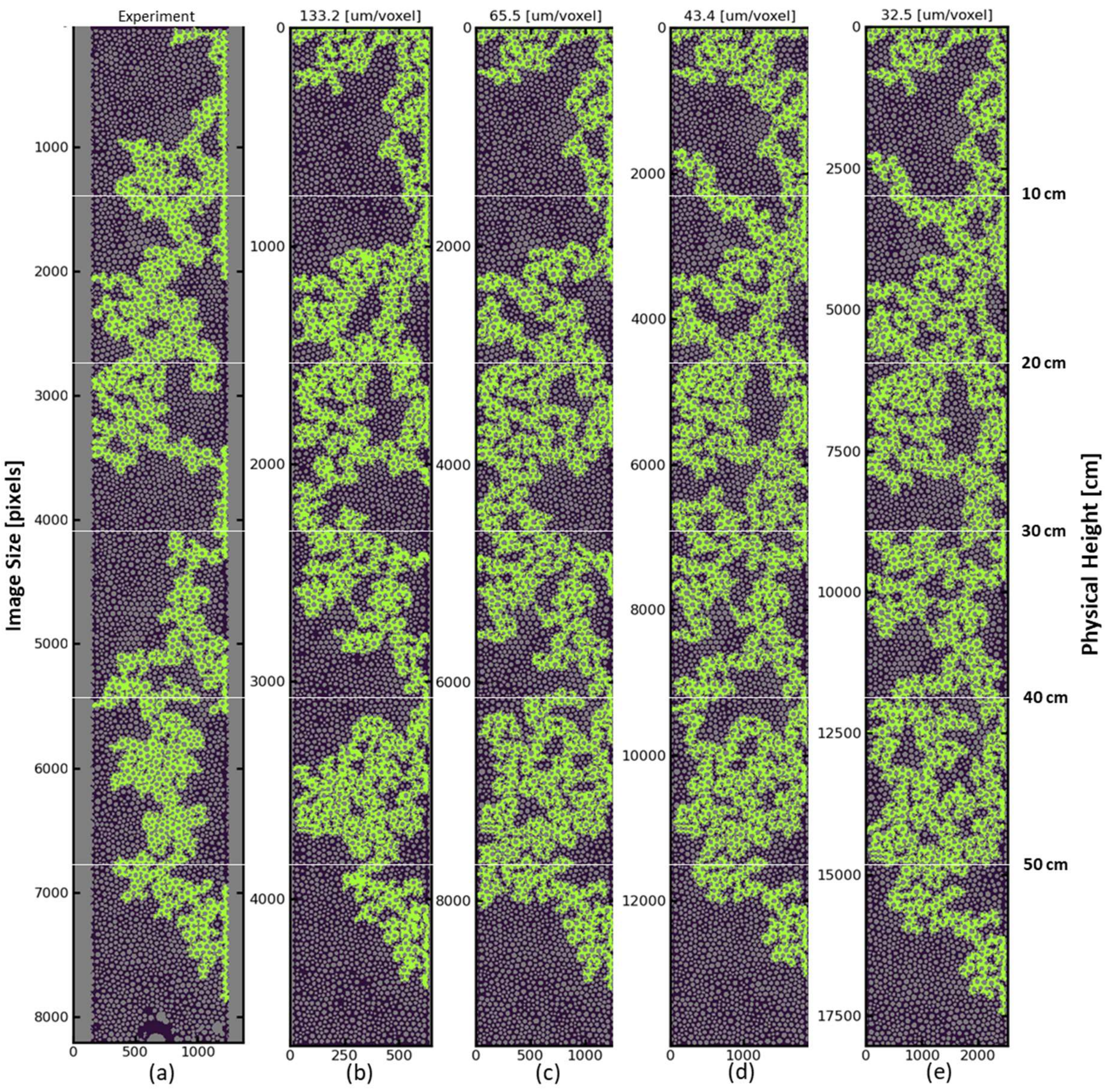

**Figure 12: Invasion patterns in micromodel as a function of image resolution. The left panel shows the experimental result, while the remaining panels show the results of the simulation on images with increasing resolution as indicated. Green pixels are in the invading non-wetting fluid, black is uninvaded (or trapped) defending phase, and grey is solid)**

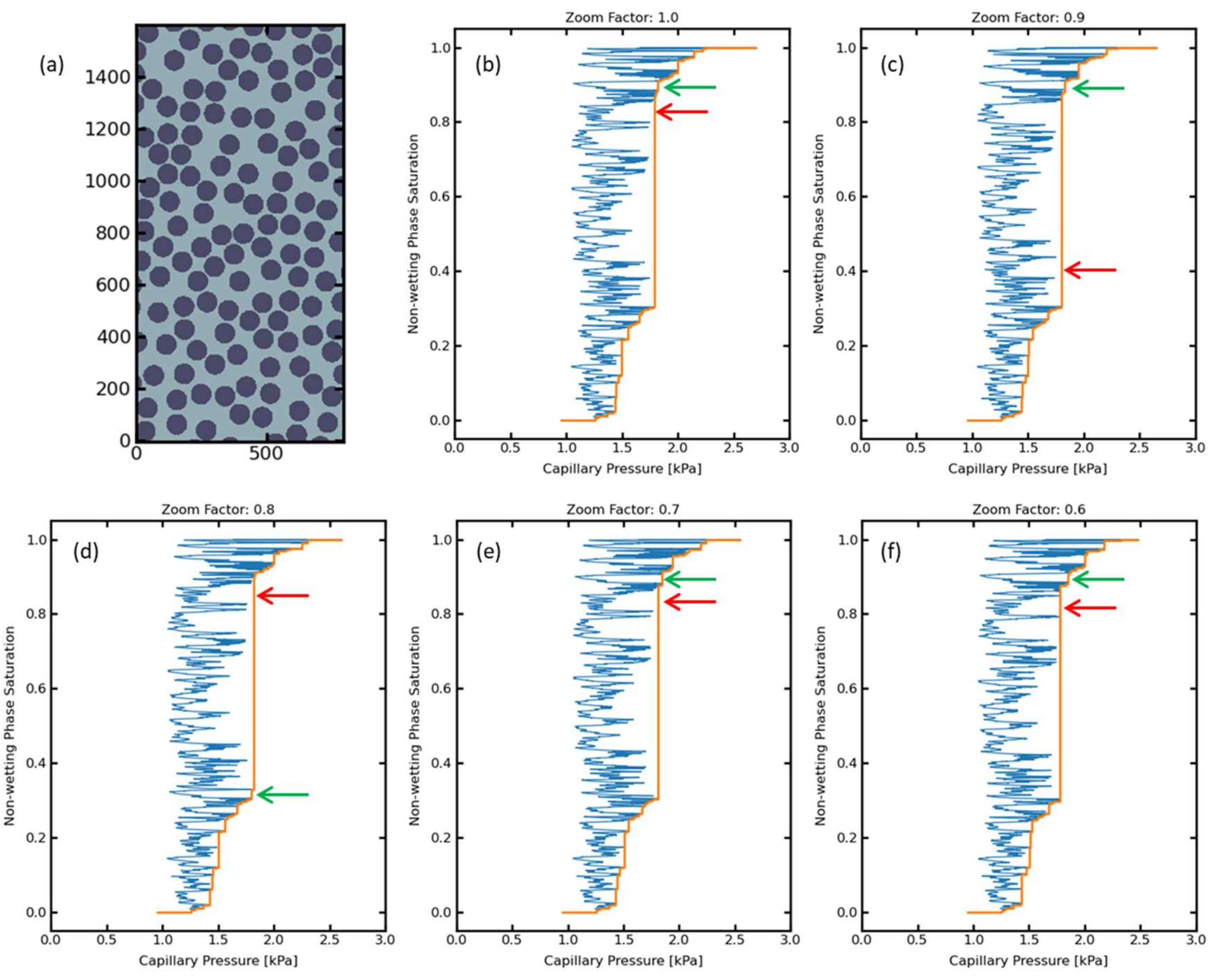

**Figure 13: Capillary pressure curves for invasion into a 2D image of randomly located non-overlapping spheres. From left to right the image resolution was reduced.**

# 4. Conclusion

The recently reported image-based invasion percolation (IBIP) algorithm [19] was extended to include gravitational effects. Moreover, a much faster implementation, termed queue-based invasion percolation (QBIP) was described which showed more than an order of magnitude improvement in processing time. Moreover, this speedup appeared consistent over the range of image sizes, which was not the case with the GPU implementation discussed previously. The IBIP and QBIP algorithms were shown to produce identical results for given conditions. In addition to a more rapid invasion simulations, an algorithm for identifying trapped pores was also presented based on the same priority queue approach and offering a similar speedup.

Without this algorithm, the utility of the QBIP algorithm would have been hampered due to the unreasonably long times required to identify trapped voxels using the previous cluster-based method. Again, the queue-based and cluster-based methods were shown to produce equivalent results.

The QBIP algorithm was compared to the drainage algorithm presented by Chadwick et al [29], herein termed IBOP for image-based ordinary percolation, which could optionally incorporate gravity. It was shown that the QBIP algorithm could successfully match the output of the IBOP algorithm with and without gravity effects. The final validation was provided by comparing the QBIP simulation to micromodel gravity drainage experiments. Although a perfect match of the invasion patterns was not obtained, it was shown that the qualitative agreement was satisfactory. Quantitatively it was confirmed that the height of the non-wetting front as a function of global saturation agreed quite well.

Finally, it was noted that the invasion patterns produced by both IBIP and QBIP algorithms are sensitive to the image resolution. By their nature, invasion percolation processes are history-dependent since they advance sequentially rather than simultaneously. A divergence early on in the invasion process due to a minor image resolution error can lead to drastically different invasion pathways later on. In other words, small changes in the image resolution can lead to not only different ordering of pore filling, but to substantially different invasion patterns. However, the QBIP algorithm produces accurate quantitative results when it comes to predicting the capillary pressure-saturation curves, as well as the macroscopic fluid-fluid distributions. This limitation is certainly worth deeper investigation to determine the conditions under which it is more likely to occur and to what extent. For instance, this effect might be more pronounced in samples with relatively narrow throat sizes like the micromodel used here. Furthermore, one can take solace in the fact that a given micromodel is also known to fill differently between experiments.

Overall, the presented QBIP algorithm provides a useful way to approximate non-wetting phase invasion into volumetric images, including the effect of gravity. This algorithm is not intended to be a physically rigorous simulation but given its outstanding computational efficiency it can be considered a reasonable approximation for more computationally intensive simulation methods such as lattice-Boltzmann. Finally, all the algorithms described here are included in the open source package PoreSpy [30].

## Acknowledgement

J.T.G. and N.M. would like to thank CANARIE (RS3-141) and the Natural Science and Engineering Research Council of Canada (RGPIN-2023-03741) for providing financial support for this project. J.T.G. would like to gratefully acknowledge the support of the Azzam-Dullien Professorship in Transport in Porous Media. A.I. and B.Z. would like to thank the NSERC Discovery Grants (RGPIN-2019-07162) for providing financial support for this project.

## Availability Statement

The QBIP, IBIP and IBOP algorithms used for simulations in this work are included in the PoreSpy package, which is preserved at DOI: 10.5281/zenodo.2633284, available via MIT License and developed openly at https://github.com/PMEAL/porespy.